# An active filament on a cylindrical surface: morphologies and dynamics

Chen Shen, Chao-ran Qin, Tian-liang Xu, Kang Chen*, Wen-de Tian*

1. Center for Soft Condensed Matter Physics & Interdisciplinary Research, School of Physical Science and Technology, Soochow University, Suzhou 215006, China

Email: tianwende@suda.edu.cn, kangchen@suda.edu.cn

**Abstract:** Structure and dynamics of an active polymer on a smooth cylindrical surface are studied by Brownian dynamics simulations. The effect of active force on the polymer adsorption behavior and the combined effect of chain mobility, length $N$, rigidity $\kappa$, and cylinder radius, $R$, on phase diagrams are systemically investigated. We find that complete adsorption is replaced by irregular alternative adsorption/desorption process at a large driving force. Three typical (spiral, helix-like, rod-like) conformations of the active polymer are observed, dependent on $N$, $\kappa$, and $R$. Dynamically, the polymer shows rotational motion in spiral state, snake-like motion in the intermediate state, and straight translational motion without turning back in the rod-like state. In the spiral state, we find that rotation velocity $\omega$ and chain length follows a power-law relation $\omega \sim N^{-0.42}$, consistent with the torque-balance theory of general Archimedean spirals. And the polymer shows super-diffusive behavior along the cylinder at long time in the helix-like and rod-like states. Our results highlight the mobility, rigidity, as well as curvature of surface can be used to regulate the polymer behavior.

## 1. Introduction

Polymer behavior on a cylindrical surface is a fundamental issue in polymer physics, which is closely related to some important biological processes and technological applications. A typical example is the winding of DNA around histone octamers, which is important for chromatin formation.[1] Another example is the adsorption of polymers on the carbon nanotube surface, which is important for multifunctional applications.[2] A polymer adsorbing on a nanotube in the thermodynamic equilibrium state has been intensely studied by theories[3] and simulations[4,5]. For instance, the compact spherical droplets, crescent-shape, and barrel-like conformations of a flexible polymer were obtained by Vogel et al[6] via Monte Carlo techniques. The polymorphism of polymer conformations arises from the competition of entropy, adsorption energy, and bending energy. Three types of local conformations (i.e., non-helical loop, helical wrapping, and straight extension) of a semi-flexible polymer with different chain length and stiffness were reported by Guo et al[7] because of high elastic energy penalty and conformation transition hindered by long chain.

Naturally, living polymers are in a non-equilibrium state[8]. For example, chromosome segregation is regulated by microtubule, which imposes a pulling/pushing force on the chromosomal loci.[9–11] Besides, DNA behavior is tuned by bio-enzymes such as DNA polymerase, helicases,[12] which can convert chemical energy (ATP) into mechanical energy and hence exert an internal stress/tension to DNA.[13] Moreover, microfilament, which is regarded as thin and long polymer chains, shows a tread-milling motion in the ATP solution[14] or associating with motor proteins.[15–17] To understand how active force governs the non-equilibrium dynamics of biopolymers, an active polymer model is proposed recently,[18,19] where a self-propulsion force is imposed tangentially to the polymer. The active polymer displays a rich phenomenology[20,21] such as super-diffusion, rotational motion, helical motion[22,23], and collective motion[24]. Although the structure and dynamics of active polymers in the bulk and two dimensions have drawn immense interest in recent years[18,25,26], their behavior near the surface such as bio-membrane and tube-like substrates has been barely investigated[27–29]. The motivation to do so is twofold: first, from a fundamental point of view, there is a need to know how the active force changes the behavior of a polymer at interface, which should be different from that of polymer in the equilibrium state. It is more interesting for polymer physicists to understand non-equilibrium polymer systems. Second, from practical consideration, polymer chains are ubiquitous in biological systems, in addition to their wide industrial applications. The biological systems are inherently non-equilibrium and the interface is very rich due to massive organelles in these systems. Understanding the interface behavior of non-equilibrium polymer is helpful to design the self-healing biomaterials.

In this article, we focus on the behavior of an active polymer on a cylindrical surface using Langevin dynamics simulations[27]. In particular, we pay attention to cooperative effects of active force, stiffness, chain length, and surface curvature on the polymer conformation. We find three typical conformations, namely, spiral state, helix-like state, and straight (rod-like) state on the dependence of chain rigidity. The polymer acquires rotational motion in spiral state, snake-like motion in the intermediate state, and straight translational motion without turning back in the rod-like state. A power-law relation between rotation speed $\omega$ and chain length, $\omega \sim N^{-0.42}$, is found in the spiral state, consistent with the torque-balance theory of general Archimedean spirals. And the polymer shows super-diffusive behavior along the cylinder at long time in the helix-like and rod-like states.

The article is organized as follows. We first introduce our model and simulation methods in section 2. Then we present the main results in section 3 and explore the active effect on the adsorption-desorption behavior of the polymer. Finally, we give a conclusion in the section 4.

## 2. Model and methods

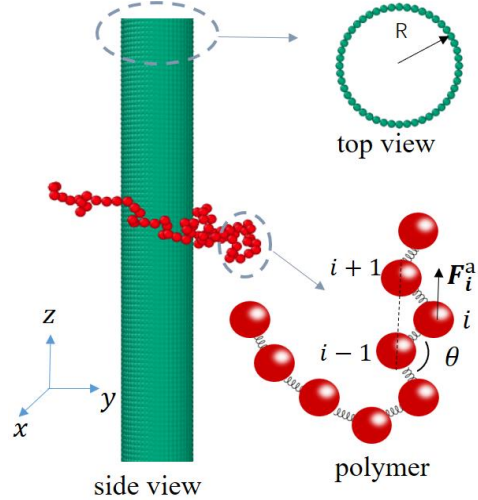

Fig.1: Schematic of the system showing a polymer near an infinite cylinder. The cylinder of radius, $R$, is built by Lennard-Jones (LJ) beads with a distance of $0.5\sigma$. The bead-spring model is adopted for the polymer with an active force, $\boldsymbol{F}_i^a$, tangential to the polymer chain. The axis of cylinder is along the Z-direction.

We consider a three-dimensional (3D) system where a polymer chain is adsorbed on an infinite cylinder which is placed along the Z-direction (Fig.1). The cylinder of radius, $R$, is built via Lennard-Jones (LJ) beads with a distance of $0.5\sigma$. The polymer consists of a sequence of $N$ monomeric units of mass, $m$, connected by harmonic springs. The monomers and cylindrical beads interact with each other through an attractive LJ potential:

$$U_{LJ} = \sum_{i=1}^{N}\sum_{j=1} u_{LJ}(\boldsymbol{r}_{i,j})$$

$$u_{LJ}(r_{i,j}) = \begin{cases} 4\varepsilon\left[\left(\frac{\sigma}{r_{i,j}}\right)^{12} - \left(\frac{\sigma}{r_{i,j}}\right)^{6}\right] & \text{for } r_{i,j} < 2.5\sigma \\ 0 & \text{for } r_{i,j} \geq 2.5\sigma \end{cases} \quad (1)$$

where $r_{ij}$ is the distance between the $i$th monomer of active polymer and the $j$th cylindrical bead, $\varepsilon$ the well depth and $\sigma$ the diameter of monomers or beads.

The internal potential energy of monomers in the polymer includes three contributions:

$$U_{intra} = \sum_{i=1}^{N-1}\sum_{j>i}^{N} U_{WCA}(r_{i,j}) + U_{bond} + U_{angle} \qquad (2)$$

where $U_{WCA}(r_{ij}) = 4\varepsilon\left[\left(\frac{\sigma}{r_{ij}}\right)^{12} - \left(\frac{\sigma}{r_{ij}}\right)^{6}\right] + \varepsilon$ for $r_{ij} < 2^{1/6}\sigma$, and $U_{WCA}(r_{ij}) = 0$ for $r_{ij} \geq 2^{1/6}\sigma$, which is the excluded volume interaction between all monomers in the polymer. The spring potential is given as $U_{bond} = k\sum_{i=1}^{N-1}(|r_{i,i+1}| - r_0)^2$ with the spring constant $k = 10000\varepsilon/\sigma^2$ and the equilibrium bond length $r_0 = 1.0\sigma$. The bending potential $U_{angle} = \kappa\sum_{i=1}^{N-2}(\theta_{i,i+1} - 180°)^2$ accounts for the stiffness of the polymer. Here $\kappa$ is the bending rigidity, $\theta_{i,i+1}$ is the bending angle of the $i$th bond and the ($i+1$)th bond.

The motion of monomers is described by the Langevin equations:

$$m\ddot{r}_i = -\nabla_i U - \gamma \dot{r}_i + F_i^r + F_i^a \qquad (3)$$

where $r_i$ is the coordinates of monomer $i$, $\gamma$ the friction coefficient, $U = U_{LJ} + U_{intra}$ including the attractive interaction with cylindrical surface and intra-molecular interaction. $F_i^r$ is the thermal noise force, which satisfies the fluctuation-dissipation relation, $\langle F_i^r(t) \rangle = 0$, $\langle F_i^r(t) \cdot F_j^r(t') \rangle = 6k_B T\gamma \delta_{ij}\delta(t-t')$. $F_i^a = f_a t(r_i)$ is the active force imposed on each monomer except the terminals. $t(r_i) = \frac{r_{i+1} - r_{i-1}}{|r_{i+1} - r_{i-1}|}$, unit tangent vector on the $i$th monomer. The hydrodynamics interaction is ignored in the simulations.

We use the home-modified LAMMPS software to perform simulations[30]. All beads of the cylinder are immovable in the simulation. The simulation box is $80\sigma \times 80\sigma \times 100\sigma$ with periodic boundary condition in all directions, which is large enough to avoid the finite size effects. Reduced units are used in the simulation by setting m=1, $\varepsilon$=1, and $\sigma$ = 1. The corresponding unit time, $\tau = \sqrt{m\sigma^2/\varepsilon}$. We choose the reduced temperature, $k_B T = 1.2$, and set the friction coefficient, $\gamma = 10$, which results in the motion of monomers effectively overdamped. For each case, a polymer after sufficient relaxation is placed close to the cylinder, then the simulation is performed by a total time of $3\times10^4\tau$ with a time step, $dt = 0.001\tau$. Some cases are performed six independent runs for ensemble averaging. For comparison, systems without active force are also performed. The detailed method is given in Supporting Information (SI)

## 3. Results and discussion
### 3.1 Impact of the active force

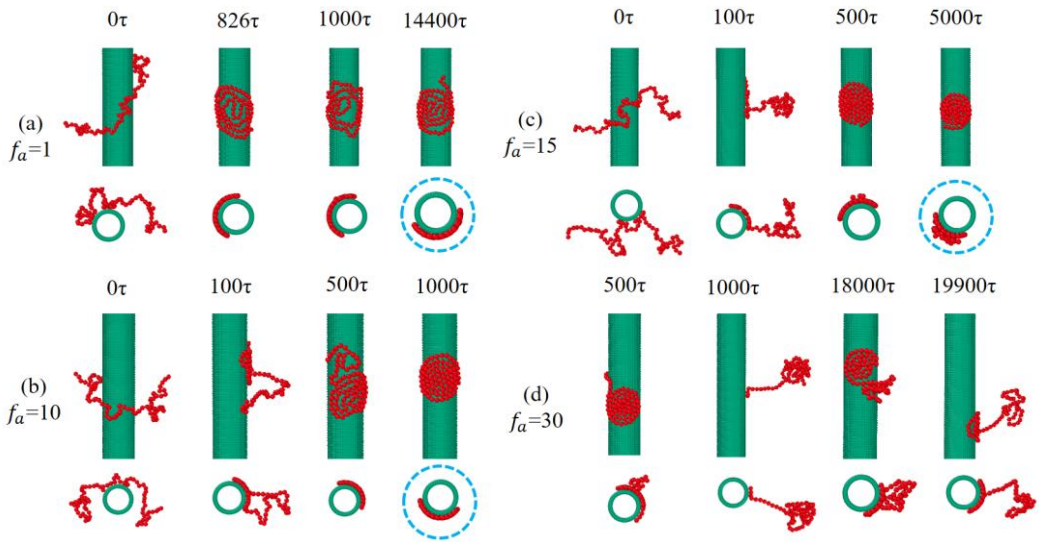

Fig.2 Typical snapshots of polymer illustrating the adsorption and desorption process for various active forces at $N$=80, $R$=3, $\kappa = 1.0$. The snapshots in blue circles demonstrate the single-layer spiral state ($f_a$=1.0, 10) and a double-layer spiral state ($f_a$=15).

We first consider a very flexible polymer with $N = 80$, $\kappa = 1.0$ near the cylinder of radius $R = 3$ and pay particular attention to the active induced adsorption/desorption behavior. Typical snapshots for various active forces, $f_a$s, are given in Fig.2 (as well as Fig.S1 in SI). At small active force ($f_a = 1.0$), the chain is adsorbed on the cylindrical surface in a single layer and wriggles with the elapsed time. It could spontaneously wind up to a loose spiral. However, the loose spiral is not stable and it breaks up with the time evolution. At intermediate active force ($f_a = 5.0$ and $10.0$), the polymer winds up to a tight, single-layer spiral on the surface and rotates continuously (Mov.S1). The adsorption energy, which is the total potential energy between the surface and polymer, sharply decreases and then maintains about $-6.3 \times 10^3 \epsilon$ (Fig. 3a), indicating all monomers are adsorbed on the surface. The radial distribution function of polymer monomers around the Z-axis of cylinder displays one peak around r~4σ (Fig.3b), also indicating it is a single layer. The spontaneous spiral formation is a consequence of effective two-dimensional confinement of the polymer on a cylindrical surface due to strong adsorption and propelling-motion-induced self-collision (steric effect) [31].

At $f_a = 15$, a double-layer spiral forms (see Fig.2c and Fig3b). This is because the head monomer could be squeezed out of the cylindrical surface (see Mov_S2) at the active force. On the one hand, the active force is not large enough to cause full desorption of the chain. On the other hand, the extruded monomers cannot re-enter the first layer due to steric hindrance. Hence, they stick to the first layer. To explore the mechanism, the virial pressure, $\Pi = \langle \sum_{k=1}^{n}(r_k - r_t) \cdot (-\nabla U_{WCA}(r_{kt})) \rangle$, on the head monomer ($r_t$) exerted by surrounding monomers ($r_k$) was calculated; here $n$ is the number of surrounding monomers (within the distance $|r_k - r_t| < 2^{1/6}\sigma$) and the angle brackets denote ensemble average. The pressure increases with the active force (Fig.3c), implying the enhancement of the squeezing-out effect. Further increase in active force ($f_a = 30$) leads to the desorption of the polymer, manifested by the time evolution of polymer conformation (Fig.2d). The adsorption energy changes between zero (far away from the interface) and $-6.3 \times 10^3 \epsilon$ (Fig.3d), demonstrating the irregular alternative adsorption/desorption process of the polymer[32]. The desorption is triggered similarly by the thermal fluctuation of the head monomer and the stress-induced extrusion. In this case, the active force is large enough to overcome the adsorption energy [33] and leads to complete desorption of the polymer (see Mov_S3). Due to the periodic boundary condition, the polymer could be adsorbed on the cylinder again after "swimming" in the box (See Mov_S4).

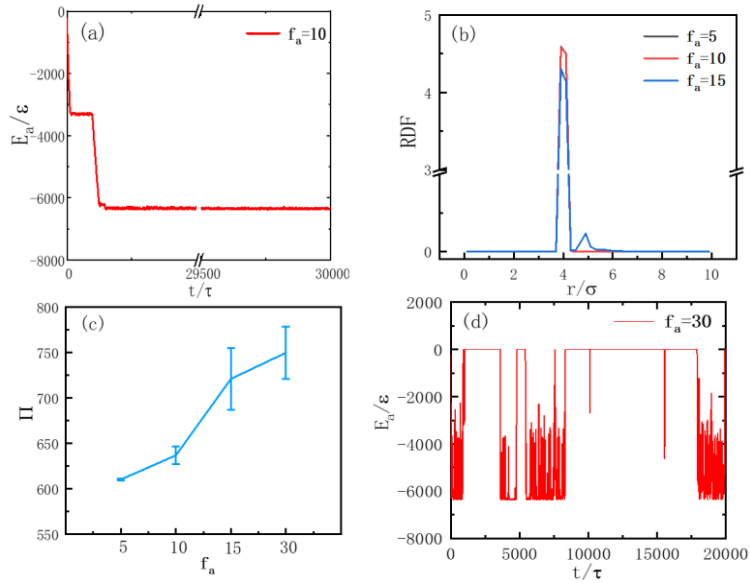

Fig.3 (a) Time evolution of adsorption energy, which is the total potential energy between the surface and polymer, at $f_a$ =10, N=80, R=3, $\kappa = 1.0$. The adsorption energy around $-6.3 \times 10^3 \epsilon$ denotes the complete adsorption state, while zero energy denotes the desorption state, where the polymer is away from the cylinder surface. (b) Radial distribution function of monomers around the cylindrical surface for $f_a$=5, 10, 15. (c) The pressure, $\Pi$, on the head monomer as a function of active force. The error bar is the standard deviation of $\Pi$. (d) Time evolution of adsorption energy between cylinder and active polymer at $f_a$ =30, $N$=80, $R$=3, $\kappa = 1.0$.

At $f_a$=10, the polymer is adsorbed on the cylindrical surface and forms a stable single-layer spiral. Below, we fix the active force $f_a$=10 so that the polymer is fully adsorbed on the cylindrical surface and explore how chain length, rigidity, and cylindrical radius influence the structure of the polymer.

### 3.2 Phase diagram

Our simulation result shows that the chain exhibits three typical conformational states: spiral state, helix-like state, and straight state. To quantitatively investigate the three states[34], we define the turning number $\psi_s = \frac{1}{2\pi}\sum_{j=1}^{s}[\arcsin(|\hat{r}_j \times \hat{r}_{j+1}|)]$, where $\hat{r}_j$ is a unit vector along the radial direction from the center of the cylinder to the $j$th bead, $s$ the index number from polymer tail to polymer head. For the straight chain, $\psi_{s=N}\sim 0$. For a chain forming an anticlockwise (clockwise) loop, $\psi_{s=N} = 1$ ($\psi_{s=N} = -1$). Then we calculate the maximum wrapping number $\alpha = \psi_s^{max} - \psi_s^{min}$ ($\psi_s^{max}$ and $\psi_s^{min}$ are the maximum and minimum of $\psi_s$ for $s$ from 1 to $N$, respectively) for each configuration. We plot the distribution function of $\alpha$ for various bending rigidities $\kappa$=10, 100, and 1000 in Fig. 4(a). It can be found that, at small $\kappa$ (= 10), the distribution of wrapping number, P($\alpha$), is very narrow with the peak at 0.35. In this case, the chain is in a stable, spiral state. At $\kappa = 100$, the $\alpha$ is distributed broadly due to the chain wrapping around the cylinder[35]. At larger $\kappa$ (=1000), the peak of P($\alpha$) is close to zero due to the rod-like configuration of the polymer along the cylinder. The average maximum wrapping number, $\langle\alpha\rangle$, as a function of $\kappa$ is given in Fig4(b). At small $\kappa(< 25)$, the chain displays as a spiral, the value of $\langle\alpha\rangle$ is close to 0.35 with a very weak deviation. As the rigidity of chain increases, spiral formation is hard due to the penalty of elastic energy. The chain wraps around the cylinder, shows a helix-like structure with a large deviation of $\alpha$ around the mean value. The $\langle\alpha\rangle$ decreases with a further increase of $\kappa$, corresponding to the chain moving like a rod along the cylinder. We use $\langle\alpha\rangle$ to quantitatively distinguish the helix-like state and the straight state. Namely, we set $\langle\alpha\rangle < \frac{1}{4}$ as the criteria of straight state if $\frac{N\sigma}{2\pi(R+\sigma)} \geq 1$, otherwise the helix-like state. For short chain ($\frac{N\sigma}{2\pi(R+\sigma)} < 1$), the criteria is $\langle\alpha\rangle < \frac{1}{4}(\frac{N\sigma}{2\pi(R+\sigma)})$ for the straight state. The spiral state is identified by viewing the trajectory and $\alpha$ distribution.

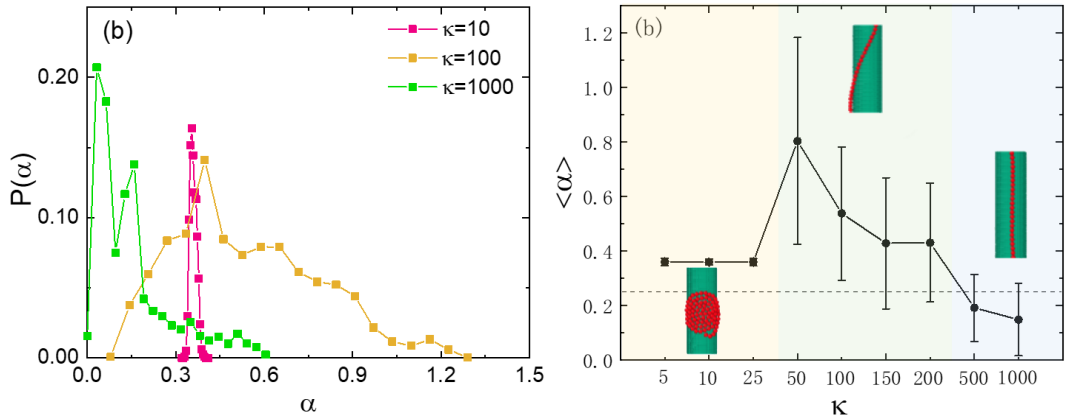

Fig.4 (a) Distribution of $\alpha$ for various bending rigidities $\kappa$ =10, 100, and 1000 at $N$=80 and $R$=3. (b) The mean maximum wrapping number, $\langle\alpha\rangle$, as a function of bending rigidities $\kappa$ ($N$=80 and $R$=3). The error bar is the standard deviation. The dashed line shows $\langle\alpha\rangle = \frac{1}{4}$, which is used as the criteria to distinguish the helix-like state and straight state.

According to the above criteria, a phase diagram in the *N*-$\kappa$ plane at $f_a = 10$ can be constructed (Fig.5a). It can be found that the spiral is easy to form at low rigidity or long chain(■). Straight configuration mainly appears at high rigidity or short chain (▲). Intuitively, the flexibility of chain is influenced by the above factors[36]. The more flexible the chain is, the easier bending the leading tip is. It is in favor of the spiral formation. Large rigidity suppresses the bending of chain, against the spiral formation. At intermediate rigidity, the polymer wraps around the cylinder (denoted by ●), like a snake motion on a tree. As a comparison, the phase diagram of polymer without active force is also given (see the background color in Fug.5 and SI). The helix-like state and straight state are denoted by the light-blue color and the light-cyan color as shown in Fig.5 (Fig.S3c also shows the typical snapshots). The polymer cannot wind up spontaneously due to the entropy effect. There are only two configurations (according to the same criteria): helix-like state and straight state. The phase boundary also shifts to the low $\kappa$ region, which implies that the active force could lower the stiffness of the semi-flexible chain.[36] Besides, instead of sliding along the cylinder in one direction, the chain randomly diffuses along the cylindrical surface due to the lack of active force.

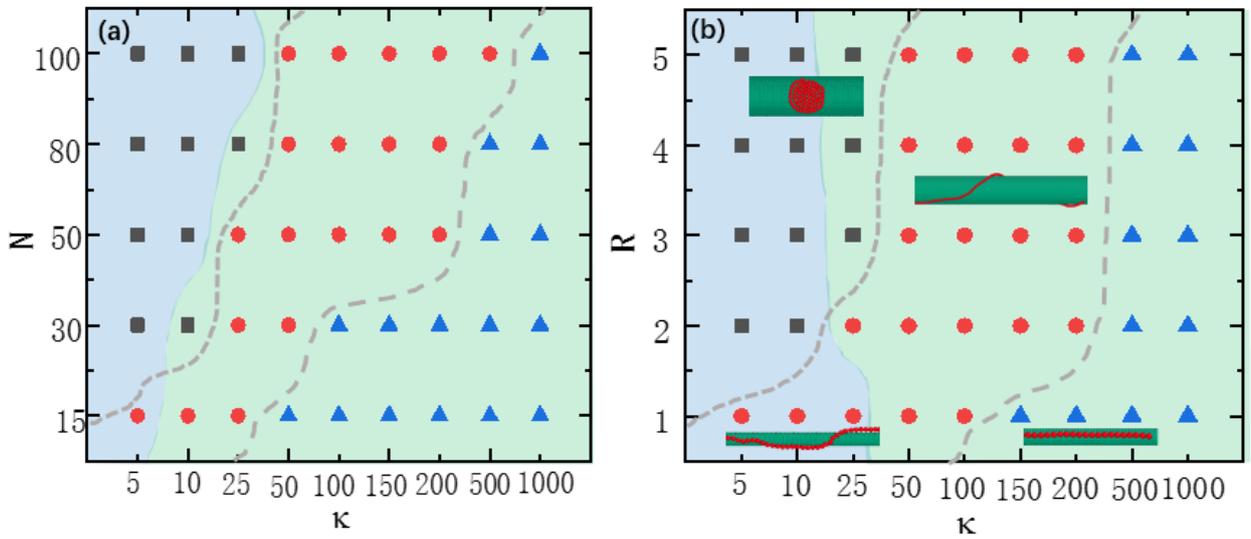

Fig.5 Phase diagram in *N*-$\kappa$ plane (a) and *R*-$\kappa$ plane (b) at *N*=80, $f_a = 10$, characterizing the spiral, helix-like, straight states. The dark square (■), red point (●), and blue triangle (▲) represent the spiral state, helix-like state, and straight state, respectively. The gray line is the eye-guided phase boundary under the active force. For comparison, the light-blue background color and the light-cyan color, respectively, denote the helix-like state and straight state without the active force. The typical snapshots are also shown for active polymer systems.

We also explore the effect of cylinder radius, *R*, and plot the phase diagram in *R*-$\kappa$ plane in Fig.5(b). Similar to Fig. 5(a), the phase diagram is roughly divided into three regions: spiral state, helix-like state, and straight state. As the bending rigidity $\kappa$ increases, the helix-like conformation will gradually translate to be a rod-like conformation[37]. Interestingly, a small *R* (high curvature of the surface) hinders the formation of spiral. The reason is that the high penalty of bending energy, which causes the instability of the spiral conformation. The bending energy originates from two aspects: one is the chain bends to take shape of a disk-like spiral, the other is the disk-like spiral deforms to fit the curvature of cylinder surface due to strong adsorption. The perimeter of small *R* smaller than the diameter of the spiral leads to the contact of spiral boundary, which can induce the broken up of spiral. Without active motion (FigS3b and FigS3d), the boundary of helix-like state and rod-like state also shifts to low $\kappa$. The phase diagram for the passive polymer at $\kappa = 25$ shows there exists a helix-like to rod-like transition with the decrease of cylinder radius. It might be caused by our criteria to distinguish the two configurations at *R*=1, which is discussed in the SI (Fig.S4).

For a similar equilibrium system, Tallury[38] explored a fully flexible polymer adheres to a zig-zag single-walled carbon nanotube (SWCNT) of fixed radius via molecular dynamics simulations. They found the flexible polymer tends to wrap around the SWCNT without any distinct conformation while a semi-flexible polymer wraps around the tube in a helical conformation. Kumar et al[39] studied the arrangement of polymer chains on a long SWCNT and they found a lamella configuration on the SWCNT of the smallest radius due to the smaller torsion energy penalty. In an equilibrium state, the conformation of polymer results from the competition between elastic energy and entropy. Elastic energy due to the bending of semi-flexible polymer drives toward a rod-like conformation while entropy favors a random configuration. Besides the active force, our model is different from these studies: we don't consider the torsion energy in our polymer model and our cylindrical surface is smooth without the chiral arrangement of beads. Thus, we observed a helix-like conformation, which is, strictly speaking, not the helical conformation found by Tallury[38] and Kumar et al[39]. It is like a state between non-helical loop[7] and helical conformation.

### 3.3 Rotation of spiral

We now turn attention to the rotation of the polymer in the spiral state. Typical snapshots of the clockwise rotation at $N=80$ $\kappa = 5, R = 3$ are shown in Fig 6a. Spiral is stable and does not break up in our finite simulation time. To quantitatively measure the rotational dynamics, we calculate the mean square angular displacement $\text{MSAD}(\Delta t) = \langle [\beta(t + \Delta t) - \beta(t)]^2 \rangle$, where $\beta(t) = \sum_0^{t-\tau} \arccos(\frac{\overline{n(t+\tau)} \cdot \overline{n(t)}}{|\overline{n(t+\tau)}| \cdot |\overline{n(t)}|})$, $\overline{n(t)}$ is the vector between two terminal monomers at the time $t$. We find the $\text{MSAD}(\Delta t)$ displays the $\sim (\Delta t)^2$ behavior for all chain lengths (Fig.6b). The rotation of the spiral could be modeled by a Langevin equation[40] as $\frac{d\beta(t)}{dt} = \omega + \xi(t)$, where $\xi(t)$ is Gaussian white noise with $\langle \xi(t) \rangle = 0$ and $\langle \xi(t)\xi(t') \rangle = 2D_r \delta(t - t')$, $D_r$ the rotational diffusion coefficient of the spiral. Integrating the equation, we obtain the $\text{MSAD}(\Delta t)$:

$$\text{MSAD}(\Delta t) = 2D_r(\Delta t) + \omega^2 (\Delta t)^2 \qquad (4)$$

where the first term corresponds to the short timescale dynamics due to thermal fluctuations and the second term corresponds to the long-time dynamics. The net rotation is observed with an angular velocity $\omega$. We use equation 4 to fit our simulation data and get the angular velocity $\omega$ for each chain length. It can be found that there exists a power-law relation between the $\omega$ and chain length $N$ with $\omega \sim N^{-0.41 \pm 0.01}$ (Fig.6b).

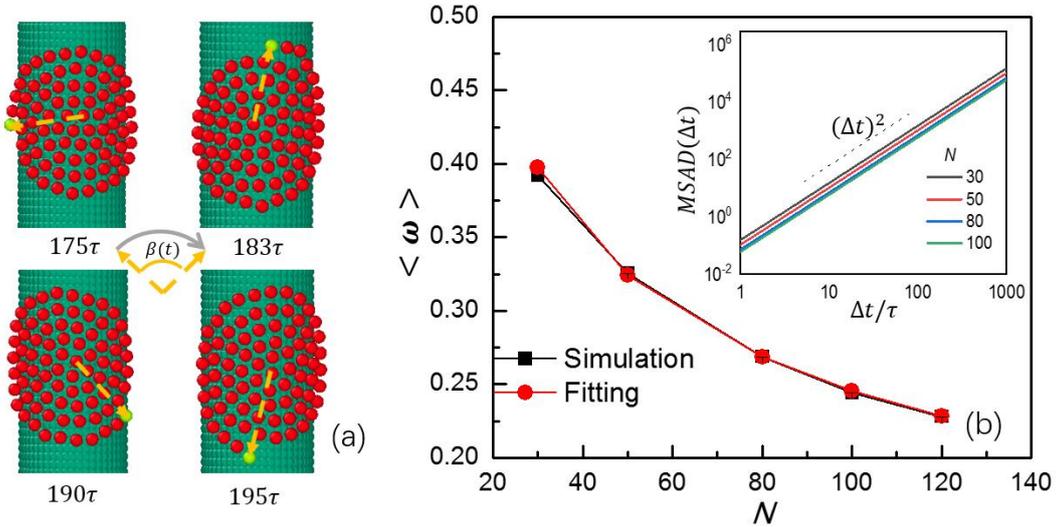

Fig.6 (a) Schematic diagram of the rotational angle $\beta(t)$. (b) Mean rotation speed $\omega$ as a function of chain length $N$ at $\kappa = 5, R = 3\sigma$. The inset is mean squared angular displacement as a function of time for $N=30,50,80,100$.

The spiral rotation can also be witnessed by the evolution of angle, $\phi(t)$, between the terminal bond vector and z-axis (as shown in Fig.7a), which displays an oscillation behavior. $\omega$ can also be calculated by Fourier transform of time evolution of the angle (the inset of Fig.7a). Using the method, we get a similar result, $\omega \sim N^{-0.42 \pm 0.01}$ (Fig.7b). To understand the physical process, we build a simple theory for the general Archimedean spiral based on the balance of the active torque, $\Gamma_1$, and drag torque, $\Gamma_2$ (see SI). In polar coordinates $(\rho, \vartheta)$, the general Archimedean spiral can be described by the equation $\rho = c\vartheta^\nu$, where c controls the distance between loops, $\nu$ the exponential factor. For normal Archimedean spiral $\nu = 1$. We get $\omega = \frac{3f_d}{(2\nu+1)\gamma c\vartheta^{\nu-1} \cdot hypergeom\left(\left[-\frac{1}{2}, \frac{3\nu}{2}\right], \left[\frac{3\nu}{2}+1\right], -\frac{\vartheta^2}{\nu^2}\right)}$, where $hypergeom()$ is a Gaussian hypergeometric function (The detail is given in SI). $N = c\vartheta^\nu hypergeom\left(\left[-\frac{1}{2}, \frac{\nu}{2}\right], \left[\frac{\nu}{2}+1\right], -\frac{\vartheta^2}{\nu^2}\right)$. We find via numerical calculation that the power-law relation between $\omega$ and $N$ exhibits for Archimedean spiral with a positive $\nu \in (0.5, 1.2)$ (Fig.S5a). For the cylindrical surface, $\omega \sim N^{-0.42}$, (Fig.7b and Fig.S3) is different from the result of two-dimension simulations, $\langle\omega\rangle \sim N^{-0.44}$ (Fig.S5b). The $\nu$ is a little bit smaller. A possible explanation is that: the decrease of drag torque resulting from the bending of a disk is more than that of active torque, which can be seen from Equation 4 and 5 in the SI, where $\Gamma_2$ is the integration of $\sim\rho^2$ while $\Gamma_1$ is the integration of $\sim\rho$. (Here $\rho = |\vec{r}|$ used in SI)

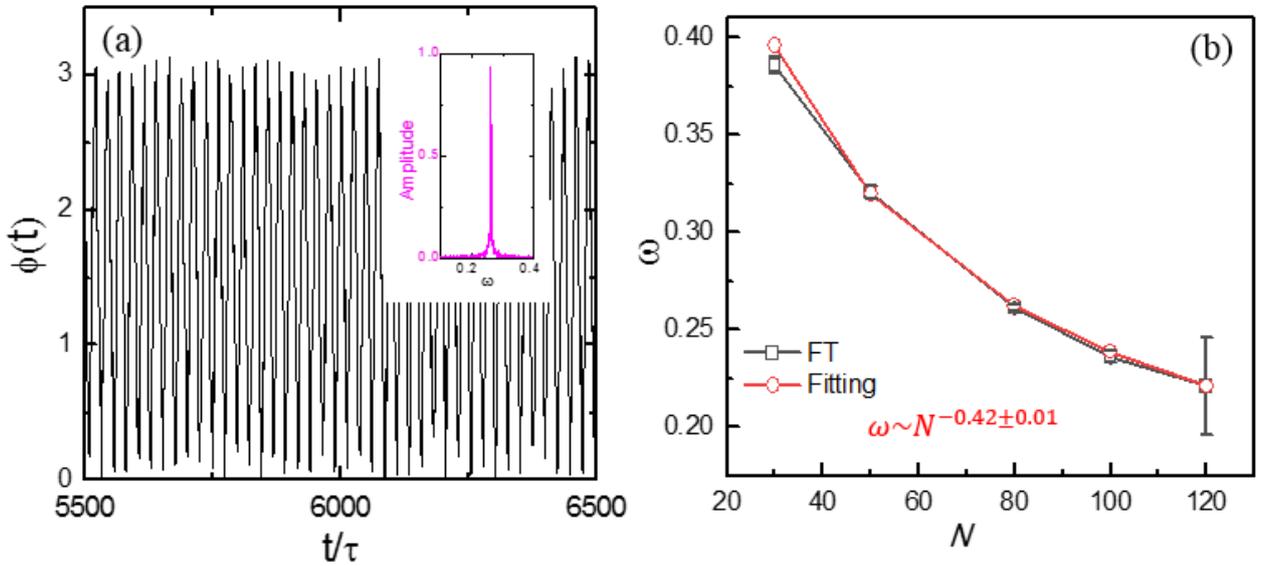

Fig.7 (a) The evolution of angle, $\phi(t)$, between the terminal bond vector and Z-axis at κ=5, R=3, N=80. The inset is Fourier transform (FT) of $\phi(t)$, showing the large amplitude occurs at $\omega \approx 0.26$. (b) Rotation speed $\omega$ by FT method as a function of $N$ at $\kappa = 5, R = 3$. $\omega \sim N^{-0.42\pm0.01}$ is witnessed.

### 3.4 Super-diffusion

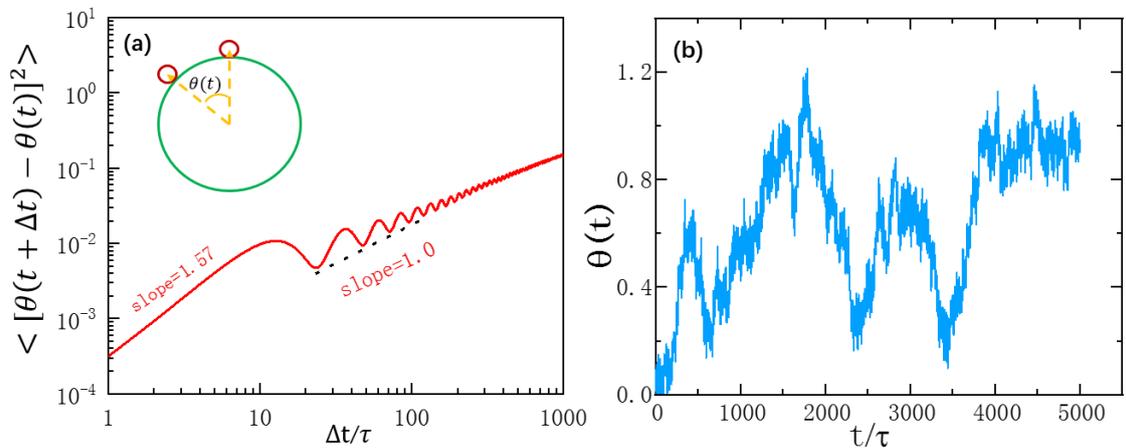

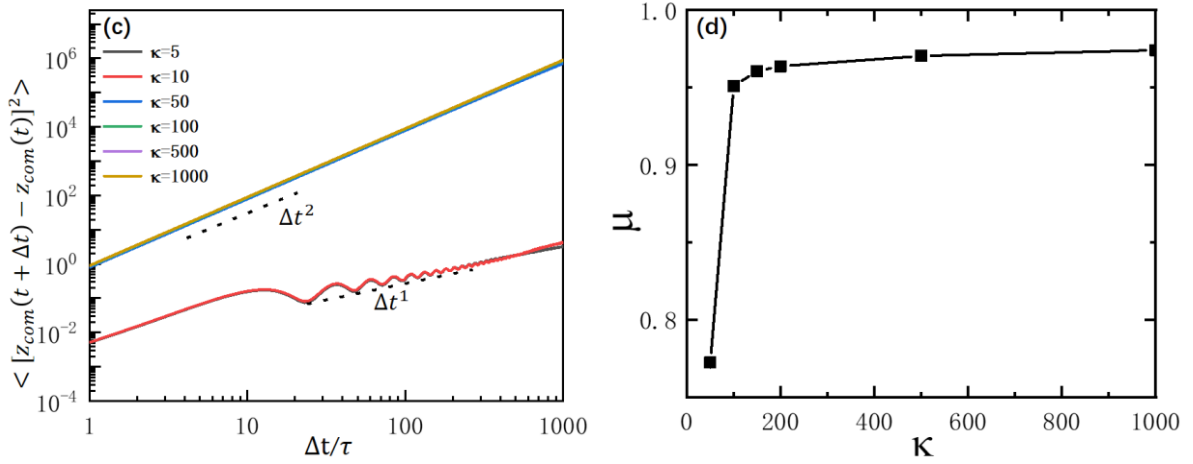

Fig.8 (a) Mean square angular displacement of angle $\theta$ for the system at $\kappa=5$, $R=3$, $N=80$, $f_a=10$. (b) The time evolution of angle $\theta$ at $\kappa=5$, $R=3$, $N=80$. The behavior for other $Ns$ is similar, which is not shown for clarification (c) The mean square displacements of the center of mass of polymer along the z axis for various $\kappa$s. (d) The fitting parameter $\mu$ for various $\kappa$s.

Then, we begin to analyze the motion behavior of the active polymer in the spiral state around cylinder from the top of view. The diffusive behavior is characterized by a mean square angular displacement $\text{MSAD}(\Delta t) = \langle [\theta(t + \Delta t) - \theta(t)]^2 \rangle$, where $\theta(t) = \sum_0^t \arcsin(\frac{|\overline{m(t+\tau)} \times \overline{m(t)}|}{|\overline{m(t+\tau)}| \cdot |\overline{m(t)}|})$, $\overline{m(t)}$ is the vector from the center of the cylinder to the center of mass of the spiral at the time $t$, as shown in the inset of Fig8a. At a short time scale, the center of mass of the spiral is super-diffusive, $\sim (\Delta t)^{1.57}$, as shown in Fig.8a. At intermediate time scale[35,41], it displays a normal diffusion, while at long time scale, it turns into a sub-diffusive state due to the fluctuation of center-of-mass of the spiral adsorbed on the cylinder. This can also be found from the time evolution of $\theta(t)$ (Fig.8b), which shows an irregular oscillation.[42] The mean square displacement of center-of-mass, $\text{MSD}(\Delta t) = \langle [z_{com}(t + \Delta t) - z_{com}(t)]^2 \rangle$, of the polymer along the z-axis for various $\kappa$s is given in Fig8c. For the spiral state, it shows a super-diffusion at short-time scale and normal diffusion at long-time scale[37]. For helix-like state and rod-like state, the translocation behavior of the polymer is like a snake on a tree. The ballistic regime presents in the time scale of our simulation with $\text{MSD}(\Delta t) \propto \Delta t^2$. We extract the effective velocity, $\upsilon$, using the method like angular velocity. The effective velocity can be predicted via $\upsilon = \frac{\mu(\kappa)(N-2)f_a}{N\gamma}$ as a balance of the net active force and total friction force. $\mu(\kappa)$ is the fitting parameter, which is a function of polymer rigidity. Intuitively, for a rod moving along the z-axis, $\mu(\infty) \sim 1$, $\upsilon$ has the maximal value[43]. This is consistent with our result: Fig8d shows $\mu$ is close to 1 at large $\kappa$. As the rigidity decreases, the leading tip swings with a larger amplitude, which cause the decrease of $\mu(\kappa)$.

For helix-like state($\kappa=50$), a large fluctuation of leading tip due to thermal effect results in the turning back of the polymer (Fig9a). This can also be seen from the time evolution of $\phi(t)$ in Fig 9b, where the $\phi(t)$ sharply drops at t~6000τ (Fig 9b and Mov_S5). Probability of turning back decreases with polymer rigidity. For rod-like state, the fluctuation of $\phi(t)$ is very small at $\kappa=500$ (Fig.9b). Turning back of the leading tip is hard to occur.

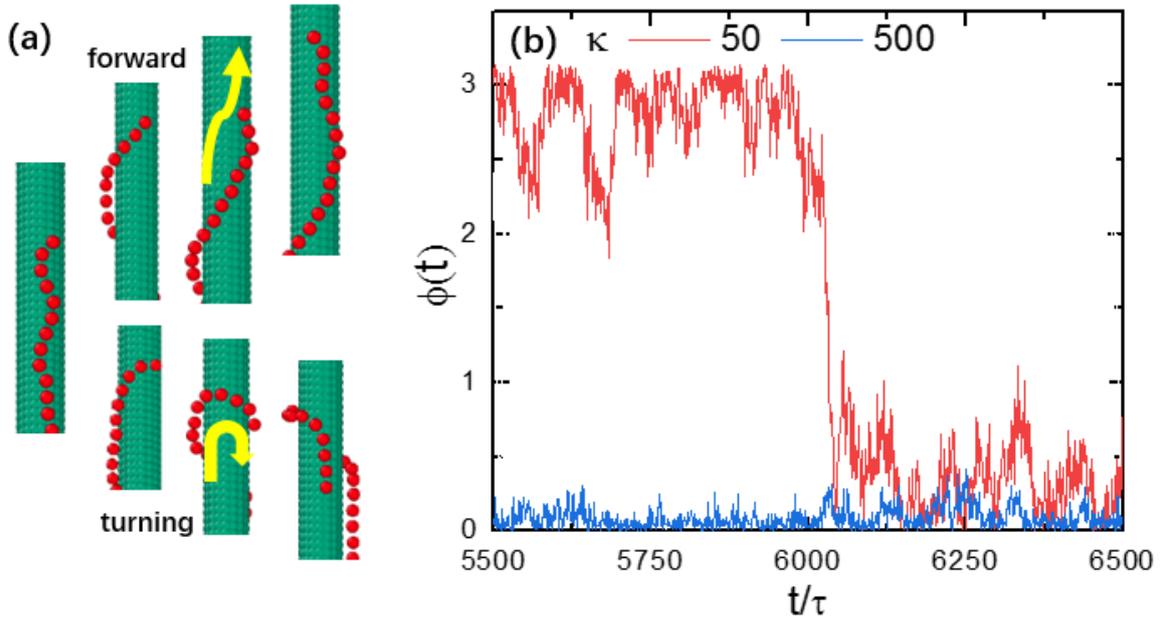

Fig.9 (a) Typical snapshots of forward and turning back of the polymer at intermediate rigidity of κ=50. (b) The evolution of angle, $\phi(t)$, between the terminal bond vector and Z-axis.

## 4. Conclusion

Using Brownian dynamics simulations, we have investigated the conformational and dynamical properties of an active polymer adsorbed on a cylindrical surface. We first studied the effect of the active force on the adsorption-desorption behavior of a flexible chain. Without driving, the polymer adsorbs on the surface in a coil state. With the increase of active force, a single-layer spiral state and a double-layer state were obtained. At large active force, the polymer could leave away from the surface due to the extrusion of leading tip. Further, the structural phase diagrams dependent on $N$—$\kappa$ and $R$—$\kappa$ show three typical conformations: a rotational spiral of flexible polymer, a helix-like conformation of semi-flexible polymer, and rod-like state of rigid polymer. Rigidity not only plays a crucial role in the conformation of the polymer but also affects its dynamics. The spiral polymer rotates at a uniform speed, $\omega$, on the power-law dependence of $N$ with $\omega \sim N^{-0.42 \pm 0.01}$. Via assuming a general Archimedean spiral, we derived an analytical expression of $\omega$ and $N$ based on torque balance between active and drag force, proving the existence of power-law relation. Finally, we found that the semi-flexible chain wraps around the cylinder and presents a super-diffusion behavior. The rigid chain is adsorbed in a straight line along the Z-axis of the cylinder and almost moves at a constant speed.

It should be noted that this is the first step to understand the active polymer adsorbed on cylinder. There is only one active polymer chain in our simulation system, which corresponds to a very dilute solution. With the increase of the polymer concentration, collective behavior of polymers might appear near the interface due to cooperative effect. Also, the hydrodynamics was neglected in our model, which will be studied in the future work.

**Acknowledgments**

This work was supported by the National Natural Science Foundation of China (NSFC). W. Tian acknowledges financial support from NSFC Grant Nos. 21674078. K. Chen also acknowledges financial support from NSFC Grant Nos.21774091, 21574096.